# Electrostatic AB-Ramjet Space Propulsion*

**Alexander Bolonkin**
*C&R, 1310 Avenue R, #6-F, Brooklyn, NY 11229, USA. aBolonkin@gmail.com <aBolonkin@juno.com>,
<http://Bolonkin.narod.ru>*

**Abstract**

A new electrostatic ramjet space engine is proposed and analyzed. The upper atmosphere (85 -1000 km) is extremely dense in ions (millions per cubic cm). The interplanetary medium contains positive protons from the solar wind. A charged ball collects the ions (protons) from the surrounding area and a special electric engine accelerates the ions to achieve thrust or decelerates the ions to achieve drag. The thrust may have a magnitude of several Newtons. If the ions are decelerated, the engine produces a drag and generates electrical energy. The theory of the new engine is developed. It is shown that the proposed engine driven by a solar battery (or other energy source) can not only support satellites in their orbit for a very long time but can also work as a launcher of space apparatus. The latter capability includes launch to high orbit, to the Moon, to far space, or to the Earth's atmosphere (as a return thruster for space apparatus or as a killer of space debris). The proposed ramjet is very useful in interplanetary trips to far planets because it can simultaneously produce thrust or drag and large electric energy using the solar wind. Two scenarios, launch into the upper Earth atmosphere and an interplanetary trip, are simulated and the results illustrate the excellent possibilities of the new concept.

**Key words**: Electrostatic ramjet space engine, ramjet space thruster, orbit launcher. Earth orbit space propulsion. Interplanetary propulsion, interstellar propulsion, AB-Ramjet space engine.
*Presented as paper AIAA-2006-6173 to AIAA/AAS Astrodynamics Specialist Conference, 21-24 August 2006, USA.

## Introduction

**General**. At present, we use only one method of launch for extra-planetary flight that being liquid-fuel or solid-fuel rockets. This method is very complex, expensive, and dangerous.

The current method of flight has reached the peak of its development. In the last 30 years it has not allowed cheap delivery of loads to space nor made tourist trips to the cosmos, or even to the upper atmosphere, affordable. Space flights are very expensive and not conceivable for the average person. The main method used for electrical energy separation is photomontage cells. Such solar cells are expensive and have low energy efficiency.

The aviation, space, and energy industries need revolutionary ideas which will significantly improve the capability of future air and space vehicles. The author has offered a series of new ideas[1-60] contained in a) numerous patent applications,[3-18] b) manuscripts that have been presented at the World Space Congress (WSC)-1992, 1994[19-22], the WSC-2002[23-32], and numerous Propulsion Conferences,[32-38] and c) other articles.[39-59]

In this article a revolutionary method and implementations for future space flights are proposed. The method uses a highly charged open ball made from thin film which collects space particles (protons) from a large area. The proposed propulsion system creates several Newtons of thrust and accelerates space apparatus to high speeds.

**History.** The author started closed research in this area as far back as 1965[1-2]. A series of patent applications[3-18] submitted during 1982 -1983 documented several methods and implementations for space propulsion and electric generators using solar wind and space particles. In 1987 these ideas were described in Report ESTI[16]. In 1990 the author published brief information about this topic[17] (see pp. 67 - 80) and in 1992 -1994 he reported on further research at the World Space Congresses - 1992, 1994, 2002[19-32] and his manuscripts [33-59].

**Brief information about space particles and space environment**. In Earth's atmosphere at altitudes between 200 - 400 km (fig. 8), the concentration of ions reaches several million per cubic cm. In the interplanetary medium at Earth orbit, the concentration of protons from the Solar Wind reaches 3 - 70 particles per cubic cm. In an interstellar medium the average concentration of protons is about one particle in 1 cm$^3$, but in the space zones HII (planetary nebulas), which occupy about 5% of interstellar space, the average particle density may be 10$^{-20}$ g/cm$^3$.

If we can collect these space particles from a large area, accelerate and brake them, we can get the high speed and braking of space apparatus and to generate energy,. The author is suggesting the method of collection and implementations of it for propulsion and braking systems and electric generators. He developed the initial theory of these systems.

## Short Description of the Implementation

A **Primary Ramjet** propulsion engine is shown in fig.1. Such an engine can work in one charge environment. For example, the surrounding region of space medium contains the positive charge particles (protons, ions). The engine has two plates 1, 2, and a source of electric voltage and energy (storage) 3. The plates are made from a thin dielectric film covered by a conducting layer. As the plates may be a net. The source can create an electric voltage $U$ and electric field (electric intensity $E$) between the plates. One also can collect the electric energy from plate as an accumulator.

The engine works in the following way. Apparatus are moving (in left direction) with velocity $V$ (or particles 4 are moving in right direction). If voltage $U$ is applied to the plates, it is well-known that main electric field is only between plates. If the particles are charged positive (protons, positive ions) and the first and second plate are charged positive and negative, respectively, then the particles are accelerated between the plates and achieve the additional velocity $v > 0$. The total velocity will be $V+v$ behind the engine (fig.1a). This means that the apparatus will have thrust $T > 0$ and spend electric energy $W < 0$ (bias, displacement current). If the voltage $U = 0$, then $v = 0$, $T = 0$, and $W = 0$ (fig.1b).

If the first and second plates are charged negative and positive, respectively, the voltage changes sign Assume the velocity $v$ is satisfying $-V < v < 0$. Thus the particles will be braked and the engine (apparatus) will have drag and will also be braked. The engine transfers braked vehicle energy into electric (bias, displacement) current. That energy can be collected and used. Note that velocity $v$ cannot equal $-V$. If $v$ were equal to $-V$, that would mean that the apparatus collected positive particles, accumulated a big positive charge and then repelled the positive charged particles.

If the voltage is enough high, the brake is the highest (fig.1d). Maximum braking is achieved when $v = -2V$ ($T < 0$, $W = 0$). Note, the $v$ cannot be more then $-2V$, because it is full reflected speed.

**AB-Ramjet engine**. The suggested Ramjet is different from the primary ramjet. The suggested ramjet has specific electrostatic collector 5 (fig. 2a,c,d,e,f,g). Other authors said the idea of space matter collection. But they did not give the principal design of collector. Their electrostatic collector cannot work. Really, for charging of collector we must move away from apparatus the charges. The charged collector attracts the same amount of the charged particles (charged protons, ions, electrons) from space medium. They discharged collector. All your work will be idle. That cannot work.

The electrostatic collector cannot adsorb a matter (as offered some inventors) because it can adsorb ONLY opposed charges particles, which will be discharged the initial charge of collector. Physic law of conservation of charges does not allow to change charges of particles.

The suggested collector and ramjet engine have a special design (thin film, net, special form of charge collector, particle accelerator). The collector/engine passes the charged particles ACROSS (through) the installation and changes their energy (speed), deflecting and focusing them. That is why we refer to this engine as the ***AB-Ramjet engine***. It can create thrust or drag, extract energy from the kinetic energy of particles or convert the apparatus' kinetic energy into electric energy, and deflect and focus the particle beam. The collector creates a local environment in space because it deletes (repeals) the same charged particles (electrons) from apparatus and allows the Ramjet to work when the apparatus speed is close to zero. The author developed the theory of the electrostatic collector and published it in [54]. The conventional electric engine cannot work in usual plasma without the main part of the AB-engine - the special pervious electrostatic collector.

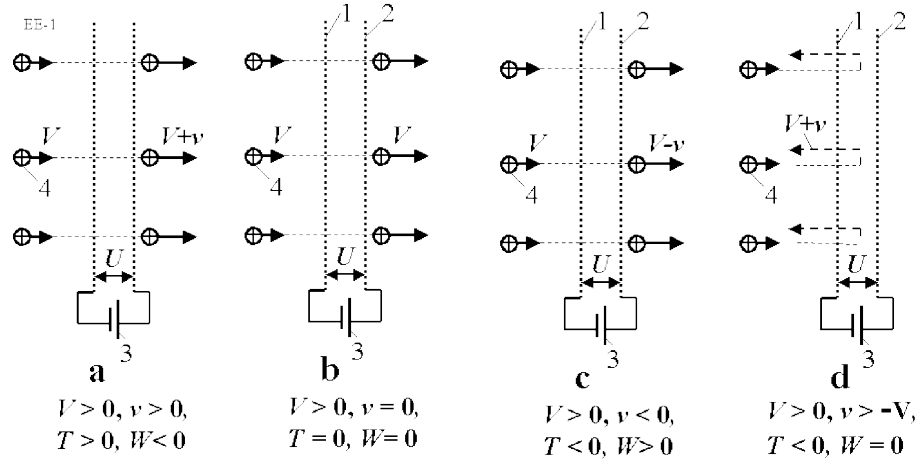

**Fig. 1**. Explanation of primary Space Ramjet propulsion (engine) and electric generator (in braking), a) Work in regime *thrust*; b) Idle; c) Work in regime *brake*. d) Work in regime *strong brake (full reflection)*. Notation: 1, 2 - plate (film, thin net) of engine; 3 - source of electric energy (voltage $U$); 4 - charged particles (protons, ions); $V$ - speed of apparatus or particles before engine (solar wind); $v$ - additional speed of particles into engine plates; $T$ - thrust of engine; $W$ - energy (if $W < 0$ we spend energy) .

The plates of the suggested engine are different from the primary engine. They have a concentrically septa (partitions) which create additional radial electric fields (electric intensity) (fig. 2b). They straighten, deflect and focus the particle beams and improve the efficiency coefficient of the engine.

The central charge can have a different form (core) and design (fig.2 c,d,e,f,g,h). It may be:
1) a sphere (fig. 2c) having a thin cover of plastic film and a very thin (some nanometers) conducting layer (aluminum), with the concentrical spheres inserted one into the other (fig.2d),
2) a net formed from thin wires (fig. 2e);
3) a cylinder (without butt-end)(fig.2f); or
4) a plate (fig.2g).

The design is chosen to produce minimum energy loss (maximum particle transparency - see section "Theory"). The safety (from discharging, emission of electrons) electric intensity in a vacuum is $10^8$ V/m for an outer conducting layer and negative charge. The electric intensity is more for an inside conducting layer and thousands of times more for positive charge.

The engine plates are attracted one to the other (see theoretical section). They can have different designs (fig.4a - 4d). In the rotating film or net design (fig.4a), the centrifugal force prevents contact between the plates. In the inflatable design (fig. 4b), the low pressure gas prevents plate contact. A third design has (inflatable) rods supporting the film or net (fig. 4c). The fourth design is an inflatable toroid which supports the distance between plates or nets (fig. 4d).

**Electric gun.** The simplest electric gun (linear particle accelerator) for charging an apparatus ball is presented in fig. 4. The design is a long tube (up 10 m) which creates a strong electric field along the tube axis (100 MV/m and more). The gun consists of the tube with electrical isolated cylindrical electrodes, ion source, microwave frequency energy source, and voltage multiplier. This electric gun can accelerate charged particles up 1000 MeV. Electrostatic lens and special conditions allow the creation of a focusing and self-focusing beam which can transfer the charge and energy long distances into space. The engine can be charged from a satellite, a space ship, the Moon, or a top atmosphere station. The beam may also be used as a particle beam weapon.

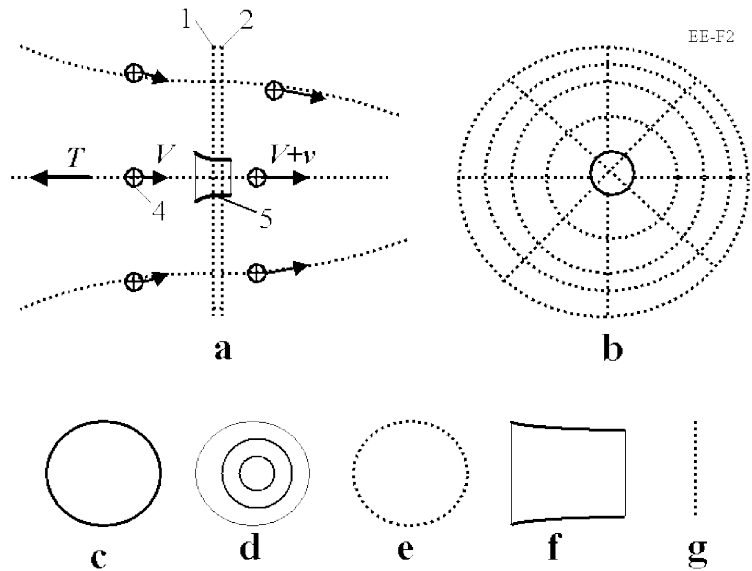

**Fig.2.** Space AB-Ramjet engine with electrostatic collector (core). a) Side view; b) Front view; c) Spherical electrostatic collector (ball); d) Concentric collector; e) cellular (net) collector; f) cylindrical collector without cover butt-ends; g) plate collector (film or net).

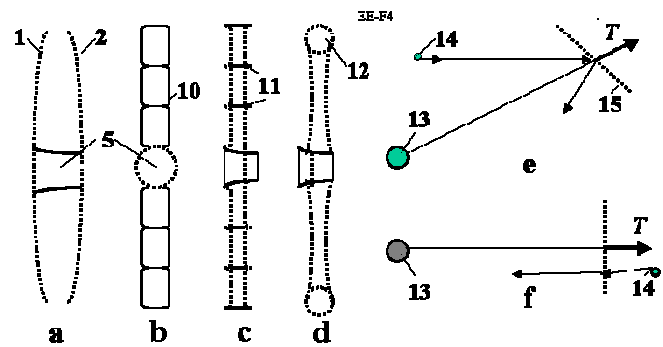

**Fig. 3.** Possible design of the main part of ramjet engine. a) Rotating engine; b) Inflatable engine (filled by gas); c) Rod engine; d) Toroidal shell engine, e) AB-Ramjet engine in brake regime, f) AB-Ramjet engine in thrust regime. Notation: 10 - film shells (fibers) for support thin film and creating a radial electric field; 11 - Rods for a support the film or net; 12 - inflatable toroid for support engine plates; 13 - space apparatus; 14 - particles; 15 - AB-Ramjet.

Approximately tens years ago, the conventional linear pipe accelerated protons up to 40 MeV with a beam divergence of $10^{-3}$ radian. However, acceleration of the multi-charged heavy ions may result in significantly `more energy.

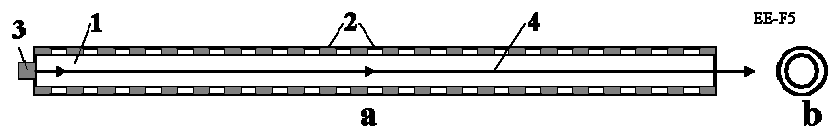

**Fig. 4.** Electric gun for charging AB-Ramjet engine and transfer charges (energy) in long distance. a) Side view, b) Front view. Notations: 1 - gun tube, 2 - opposed charged electrodes, 3 - source of charged particles (ions, electrons), 4 - particles beam.

At present, the energy gradients as steep as 200 GeV/m have been achieved over millimeter-scale distances using laser pulsers. Gradients approaching 1 GeV/m are being produced on the multi-centimeter-scale with electron-beam systems, in contrast to a limit of about 0.1 GeV/m for radio-frequency acceleration alone. Existing electron accelerators such as SLAC <http://en.wikipedia.org/wiki/SLAC> could use electron-beam afterburners to increase the intensity of their particle beams. Electron systems in general can provide tightly collimated, reliable beams while laser systems may offer more power and compactness.

## Theory of Space AB-Ramjet propulsion

The main part and innovation of the suggested system is the charged core. That may be a charged ball (fig. 2 c,d,e), a cylinder (fig. 2f), or a plate (fig. 2g). The big charge has some problems in space. We consider the major problems below.

**1. Blockading of the ball charge (AB-radius).** Blockading of the charge core by unlike particles is the main problem with this method. The charge on the core attracts unlike particles and repels like particles. The opposite charged particles accumulate near the core and block its charge. As a result, the area of ball (charged core) influence is many times less than the area of the interaction of the ball and the particles when there is no blockading. The forces are thus greatly reduced.

The author of this work proposed two models for estimation of the efficient charge radius, named AB-radius (radius of neutral working charge). That radius is analog of the Debaev shield radius for single charged particles in plasma theory. In the first model, the radius of the efficient area is computed as the area where particles of like charge to the ball are absent and the density of opposite-charged (unlike) particles is the same as the space medium. This model gives the lower limit of the efficient area. In the second model, the radius of the efficient area is computed as the area where the density of unlike particles is less than the space medium density because the unlike particles inside the efficient area have generally higher velocity than a those outside this area. The neutral area (neutral sphere) in model 2 is larger than in model 1. Model 2 is better, but this problem needs more detailed research.

*It is possible to find the minimum distance* which space electrons can approach a negatively charged ball. The full energy of a charged particle (or body) is the sum of the kinetic and potential electric energy. Any change of energy equals zero:

$$\frac{mV^2}{2} + E_p = 0, \quad E_p = \int_\infty^r F dr, \quad F = k\frac{qQ}{r^2}, \quad E_p = kqQ\left(\frac{1}{\infty} - \frac{1}{r}\right) = -\frac{kqQ}{r}, \quad \frac{mV^2}{2} - \frac{kqQ}{r} = 0 \quad (1)$$

where $m$ is the mass of a particle [kg] (mass of a proton is $m_p = 1.67 \cdot 10^{-27}$ kg, mass of an electron is $m_e = 9.11 \cdot 10^{-31}$ kg); $V$ is the speed of particle [m/s] (for solar wind $V_s = 300 - 1000$ km/s); $E_p$ is the potential energy of a charged particle in the electric field [J]; $F$ is the electric force, N; $q$ is the electrical charge of a particle [C] ($q = 1.6 \times 10^{-19}$ C for electrons and protons); $k = 9 \times 10^9$ is coefficient, $r$ is the distance from a particle to the center of the ball [m]; $Q$ is ball charge [C].

From equation (1) the minimum distance for a solar wind electron is ($m = m_e$, $V = V_s$):

$$r_{min} = \frac{2kqQ}{mV^2} = \frac{2K}{V_s^2} = \frac{2a^2 Eq}{m_e V_s^2}, \quad \text{where} \quad K = \frac{kqQ}{m}, \quad Q = \frac{a^2 E}{k}, \quad K = \frac{a^2 Eq}{m} \quad (2)$$

where $K$ is a coefficient; $m_e$ is electron mass; $V_s$ is the solar wind speed [m/s]; $a$ is the radius of ball [m]; $E$ is electrical intensity at the ball surface [V/m]. The maximum electrical intensity of the open negative bare charge is about $10^8$ - $2 \times 10^8$ V/m in a vacuum.

For $a = 6$ m, $E = 10^8$ V/m, $V_s = 4 \times 10^5$ m/s we have $r_{min} \approx 8 \times 10^6$ km.

The minimum distance of a hyperbolic particle trajectory from the punctual charged core is

$$c_h = R_e V, \quad K = \frac{a^2 qE}{m}, \quad p_h = \frac{c_h^2}{K}, \quad H \approx V^2, \quad e = \frac{c_h}{K}\sqrt{H + \frac{K^2}{c_h^2}}, \quad r_{min} = \frac{p_h}{1+e}, \quad (2a)$$

where $R_e$ is AB-radius efficiency of charged ball (see Eq. (4 - 5) later). All other values are parameters of the hyperbole and computed in (2a). The minimal radius of (2a) gives the lower estimation for the required radius of the engine plates. The above estimation of maximum plate radius for speed less 1000 km/s may be found from the equation

$$R_p = \frac{R_e^2 V^2 m}{qa^2 E}. \quad (2b)$$

2. **Minimumal neutral sphere (AB-radius) around a charged ball.**

a) **Model 1.** *Constant particle density.* The charge density of the unlike space plasma particles inside a neutral sphere is equal to the density of solar wind. The minimum radius of the neutral sphere is

$$Q = \frac{4}{3}\pi R_n^3 d, \quad R_n = \sqrt[3]{\frac{3Q}{4\pi d}} = \sqrt[3]{\frac{3a^2 E}{4\pi kd}}, \quad d = 10^6 Nq, \quad (3)$$

where $d$ is density of solar wind [C/m$^3$]; $R_n$ is the minimum radius of the neutral sphere [m]; $N$ is the number of particles in cm$^3$.

b) **Model 2.** *Variable particle density.* Density of the unlike particles inside the neutral sphere will be less than the density of solar wind particles because the particles are strongly accelerated by the ball charge to approximately the speed of light. The new density and new corrected radius can be computed in the following way:

1) The speed of protons along a ball radius is (in the system connected to the particles)

$$V_r^2(r) = V_0^2 + 2K\left(\frac{1}{r} - \frac{1}{r_0}\right), \quad (4)$$

where $V_0 = V_s$ - proton speed at an initial radius of $R >> R_n$, and $R_n$ is the radius of the neutral sphere.

2) Particle charge density, $d_p$, along a ball radius is

$$d_p = d_{po}\frac{V_0}{V_r(r)}, \quad d_{po} = 10^6 Nq/s^2, \quad s = \frac{S}{S_0}, \quad (5)$$

where $S$ is distance from the Sun in AU; $S_0 = 1$ AU; $s$ is relative distance from the Sun; $d_{po}$ is density at 1 AU.

3) Charge of the neutral sphere along a sphere radius is

$$Q_r = Q - 4\pi d_{po} V_0 \int_a^R \frac{r^2}{V_r(r)} dr. \quad (6)$$

4) The AB-radius of the neutral (blocking) sphere can be found from the condition $Q_r = 0$.

*Note.* For our estimation we can find it using the stronger condition $Q_r = 0.5Q$, and call it the efficiency radius $R_e$ of the charge $Q$. We use the stronger condition because model 2 may yield a more accurate result (with the speed $V_r$ being slower).

The efficiency radius in Model 2 is significantly more than in Model 1. Model 1 gives the lower estimation of the efficiency radius; model 2 gives the top (more realistic) estimation of the efficiency AB-radius. In our computation we will use the model 2 with the note above.

**3. Computation of main parameters of AB-Ramjet propulsion.** If we know the efficiency radius

and voltage $U$ between engine plates, we can develop the theory and compute all the main parameters of Space AB-Ramjet. The formulas and final equations are given below. All values are in metric system (SU).

1). *Additional speed v* of particles gained between engine plates.

$$\text{From} \quad \frac{mv^2}{2} = qU \quad \text{we get} \quad v = \pm\sqrt{\frac{2qU}{m}} \,, \tag{7}$$

where $m$ is mass of particle [kg](for proton $m = m_p = 1.67 \times 19^{-27}$ kg, for electron $m = m_e = 9.109 \times 10^{-31}$ kg); $q$ - charge of particle [C], for proton and electron $q = 1.6 \times 10^{-19}$ C, $U$ - electric voltage between plates [V].

2) *Mass $m_s$ running through engine* in one second

$$m_s = 10^6 SVnm \,, \tag{8}$$

where $S = \pi R_e^2$ is area of engine efficiency [m$^2$], $n$ is number particles in 1 cm$^3$ (coefficient $10^6$ in (8) transfer $n$ in m$^3$), $V$ is apparatus (or relative particles speed about apparatus, out of efficiency area (for example, Solar wind speed))[m/s].

3) *Thrust T* (or brake force, drag $D$)[N] of Ramjet engine is

$$T = m_s v, \quad T = nSV\sqrt{2mqU}, \quad D_{max} = -2m_s V \,. \tag{9}$$

The full maximum drag can be easily obtained in the conventional (fig.1) electric engine. But it is difficult to achieve in the AB-engine because collector of this engine requires very high voltage, $U_b = aE$. However, when plate voltage is zero, the AB-engine can easily achieve a slightly less than maximum drag via

$$D = c_4 m_s V \,, \tag{9a}$$

here $c_4 = 0 - 2$ is drag coefficient which depends from charged core. If $c_4$ is less 1, the drag is easily controlled using the plate voltage $U$.

4) *Currency* of particles flow through engine

$$I = nSVq \,, \tag{10}$$

5) Electric *power N* of particles flow

$$N = \pm IU \quad \text{or} \quad N_b = DV \,, \tag{11}$$

where $U$ is voltage between plates, V. This power is negative (we spend energy) when we get thrust and the power is positive when we brake, $N_b$ is brake power.

6) *Voltage* of electricity induced in brake regime

$$U_b = \frac{mv^2}{2q}, \quad v < V \,, \tag{12}$$

7) *Propulsion efficiency* coefficient.

a) For Ramjet engine the coefficient of propulsion efficiency equals

$$\eta = \frac{m_s vV}{0.5 m_s v^2 + m_s vV}, \quad \text{or} \quad \eta = \frac{1}{1 + 0.5\bar{v}}, \quad \bar{v} = \frac{v}{V} \tag{13}$$

b) For any rocket engine the coefficient of propulsion efficiency equals

$$\eta_R = \frac{TV}{TV + 0.5 m_s v^2}, \quad T = (V+v)m_s, \quad \eta_R = \frac{1}{1 + \frac{\bar{v}^2}{2(1+\bar{v})}}, \quad \bar{v} = \frac{v}{V} \,, \tag{14}$$

Computation of equations (13) - (14) are presented in fig. 5.

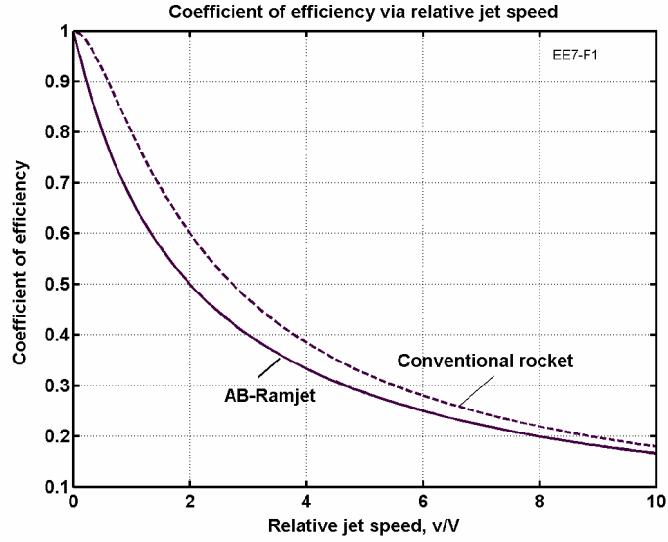

**Fig. 5.** Propulsion coefficient of efficiencies for Ramjet and conventional rocket engine (includes conventional electric rocket thrusters) versus relative jet speed.

As can be seen, the efficiency of the conventional rocket engine seems better than of the AB-Ramjet. However, full engine efficiency is a product of the propulsion and terminal efficiencies. The terminal coefficient of the conventional (liquid) rocket engine is large, with a value of 0.68 (nozzle loss), while the terminal coefficient of the conventional rocket electric (ion) thruster is small. This is because the rocket ion thruster spends a lot of energy in the ionization of the jet mass. The proposed AB-engine's terminal coefficient depends on a transparency coefficient (discussed later) of plates and core and the terminal efficiency can be high. Note that the rocket propulsion having the high speed of the jet has low efficiency from an energy viewpoint. The rocket electric thruster with high specific impulse (jet speed) spends a significant amount of energy per unit of thrust. The photon rocket has top jet speed and the worst energy efficiency. The best efficiency ($\approx 1$) is achieved by a propulsion system which repels from a very large mass (for example, a planet). Our AB-Ramjet engine uses outer space mass. That is very big advantage in comparison to the conventional electric thruster and rocket which uses its own mass.

8) Final *relative speed* of different propulsion systems. Assume we can covert mass into energy and back with an efficiency coefficient of 1. Assume the system coordinate is connected with apparatus. Compare the relative fuel consumption.

a) Photon engine.

$$, mdV = -cdm, \quad \frac{V}{c} = -\ln\left(\frac{M_k}{M_0}\right), \quad \overline{V} = -\ln \overline{M}, \quad \overline{V} = \frac{V}{c}, \quad \overline{M} = \frac{M_k}{M_0}, \quad \overline{M} = e^{-\overline{V}}, \quad \overline{M}_f = 1 - \overline{M} \qquad (15)$$

where $M_k$ is final apparatus mass [kg], $M_0$ is start apparatus mass [kg], $c$ is light speed, $c = 3 \times 10^8$ m/s, $\overline{M}_f$ is a spend relative fuel consumption.

b) Apparatus repels from a planet using its own energy (fuel) (for example, multi-reflex engine [52] located in the apparatus).

$$\text{From} \quad \frac{M_k V^2}{2} = (M_0 - M_k)c^2 \quad \text{we receive} \quad \overline{V} = \frac{V}{c} = \sqrt{2\left(\frac{M_0}{M_k} - 1\right)} = \sqrt{2\left(\frac{1}{\overline{M}} - 1\right)},$$

$$\overline{M} = \frac{1}{1 + 0.5\overline{V}^2}, \quad \overline{M}_f = 1 - \overline{M}. \qquad (16)$$

c) Apparatus repels from planet using planet energy (fuel) (for example, multi-reflex engine [52] located on planet surface).

From $\dfrac{MV^2}{2} = M_e c^2$ we receive $\overline{V} = \dfrac{V}{c} = \sqrt{2\dfrac{M_e}{M_0}} = \sqrt{2\overline{M_e}}$, $\overline{M_e} = \dfrac{1}{2}\overline{V}^2$, $\overline{M_f} = \overline{M_e}$, (17)

where $M_e$ is the mass spent on the planet surface.

The computation of the expense of mass (energy) for a perfect transfer of energy (coefficient of efficiency = 1) versus the relative apparatus speed is presented in fig. 6. Computations made in system coordinate connected with apparatus. The relativistic effect (for Earth's observer) can change these results for high speed.

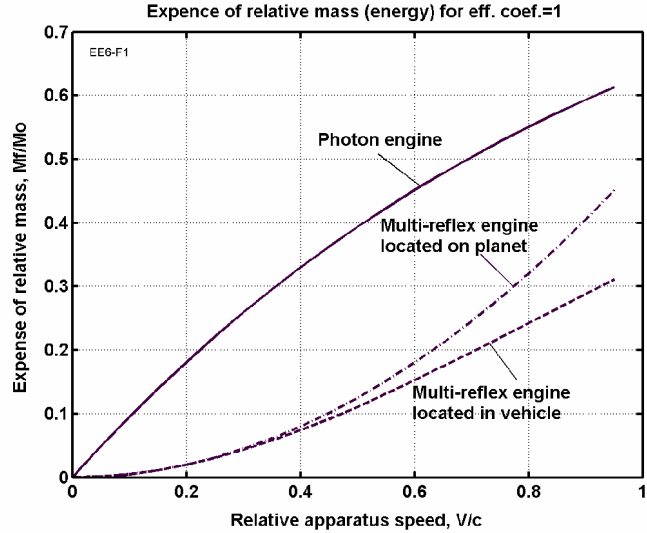

**Fig.6.** Expenses of relative mass (energy) versus relative apparatus speed for a different engines: photon engine, multi-reflex engine [52] located on the planet surface and located in the apparatus. The AB-Ramjet located between multi-reflex engine located on the vehicle and photon engine. The curve depends from a flow of space mass through engine. Transfer efficiency coefficient (mass to energy) equals 1. Computations made in system coordinate connected with apparatus. The relativistic effect (for Earth's observer) can change these results for high speed.

Note that in many ways the AB-Ramjet engine is better than the photon engine (which is a dream of all space scientists). For example, if we want to reach the relative apparatus speed of $0.1c$, the AB-Ramjet can spend 20 times less energy than the photon engine. Moreover, the AB-Ramjet engine can return spent energy (if it can transfer back self kinetic energy into mass with high efficiency). That means a space ship (using AB-Ramjet engine) can travel into space infinity time with stopping at planets (spending the mass for acceleration and return its mass in braking). But the photon engine losses part of self mass in the photon beam and can travel only limited time.

9) *Force between engine plates* is

$$f = -\dfrac{1}{2}\varepsilon_0 E^2 S_p,$$ (18)

where $\varepsilon_0 = 8.85 \times 10^{-12}$ is electrostatic coefficient [F/m], $S_p$ is plate area [m²]. For $S_p = 1$ m², $E = 1000$ V/m force $f = -4.425 \times 10^{-6}$ N.

**4. Ball (Central charged core) discharge.** The space apparatus or solar wind has high speed. This means the particles have a trajectory closed to a hyperbolic curve in the AB-area of charge influence and most of them will fly off into infinity. Only a proportion of them will travel through the ball. These particles decrease in speed and can discharge the ball. However, their speed and kinetic energy are very large because they are accelerated by the high voltage of the ball's electric field (some tens or hundreds of MV). The necessary ball film (net) is very thin (measuring only a few microns). The particles pierce through the ball. If their loss of speed is less than their (apparatus) speed or the solar wind speed, their trajectory will be close to a hyperbolic curve and they will fly into space. If their loss of speed is more than the solar wind speed, their trajectory will be close to an ellipse, so they will return to the ball and, after many revolutions, they can discharge the ball if their perigee is less then the ball's radius. This discharge may be compensated using special methods.

There are several possible methods for decreasing this discharge (Fig. 2c-g): a) A ball made of net; b), the central charge (core) has a cylindrical form (fig. 2f) without cover butt-end; c) the plates increase the particles speed to hyperbolic speed and reflect the returning particles. Note, the electric field does not depend on the form of the central charge at far distances, but does depend on the value of the charge.

Let us estimate the discharging of the plates and the central charge (ball, core).

1) If the plate or ball has a net design, the particles will cross only the wires of plate. Note, the maximum coefficient of light transparency $c_1$ is the ratio of the area of the wires, $S_w$, to the area of plate, $S_p$,

$$c_1 = S_w / S_p . \qquad (19)$$

2) The particles crossing the wire lose a part of their energy. This loss may be calculated as a brake coefficient $c_2$. The method of computation is described in the work [54].

The particle (proton) track in the matter can be computed in following way:

$$l = R_t/\gamma , \qquad (20)$$

where $l$ is track distance of the particles [cm]; $R_t = R_t(U)$ is magnitude (from a table) [g/cm$^2$]; $\gamma$ is matter density [g/cm$^3$]. The magnitude of $R_t$ depends on the kinetic energy (voltage) of the particles. For protons the values of $R_t$ are presented in Table 1.

**Table 1**. Magnitude of $R_t$ as a function of accelerated voltage $U = aE$, volts.

| U, MV | 100 | 200 | 300 | 400 | 500 | 600 | 700 | 1000 | 2000 | 3000 | 5000 |
|---|---|---|---|---|---|---|---|---|---|---|---|
| $R_t$ g/cm$^2$ | 10 | 33.3 | 65.8 | 105 | 149 | 197 | 248 | 370 | 910 | 1463 | 2543 |

The proton energy is $U = aE$. For magnitudes $a = 6$ m, $E = 10^8$ V/m, proton energy $U = 600 \times 10^6$ V and ball cover density $\gamma = 1.8$ g/cm$^3$ the proton track is $l = 197/1.8 = 109$ cm. The loss of proton energy is proportional to the wire diameter or cover thickness. Consequently, the particles brake coefficient is

$$c_2 = \frac{d_w}{l} = \frac{\delta}{l}, \qquad (21)$$

where $d_w$ is the diameter of the plate wires and $\delta$ is the thickness of the ball cover.

The full transparency coefficient $c_3$ and energy of loss $E_L$ are

$$c_3 = c_1 c_2 , \quad E_L = c_3 E , \qquad (22)$$

where $E$ is the energy of particle flow crossing the AB-Ramjet engine. Note, the energy loss across the cylindrical core (charge located on the tube) is small because particles are moved into the empty tube along its axis.

The safe ball cover thickness may be estimated using the following method. The solar wind proton energy is

$$E_d = \frac{m_p V_s^2}{2} . \qquad (23)$$

For $V_s = 400\times10^3$ km/s the proton energy is $E_d = 13.4\times10^{-17}$ J $= 13.4\times10^{-17} \times 0.625\times10^{19}$ eV $= 840$ eV.

If the loss of proton energy is proportional to the cover thickness, the maximum safe cover thickness (which will not discharge the ball) will be

$$E_d = U\frac{\delta_{max}}{l}, \quad \text{or} \quad \delta_{max} = \frac{Ed(V_s)l}{U}, \quad U = aE \quad (24)$$

For $a = 6$ m, $E = 10^8$ V/m, the required ball cover thickness is $\delta_{max} = 1.53$ micron. For $a = 4, 10$ m $\delta_{max} = 1.22$ and $1.73$ microns, respectively.

This magnitude is less than the ball thickness required for the charge stress (see [54]) for current cover matter. That way the net and cylindrical core is better then the thin-filmed spherical ball.

For electrons, the thickness of half absorption may be calculated using equation [54]

$$R_r = 0.095\frac{Z}{A}(aE)^{3/2} \quad [g/cm^2], \quad d_{0.5} = \frac{R_r}{\gamma}, \quad [cm] \quad (25)$$

Here $Z$ is the nuclear charge of the ball matter; $A$ is the mass number of the ball matter.

**5. Initial expenditure** of electrical energy needed to charge the ball. The ball must be charged with electrical energy of high voltage (millions of volts). Let us estimate the minimum energy when the charged device has 100% efficiency. This energy equals the work of moving of the ball charge to infinity, which may be computed using the equation

$$W = \frac{Q^2}{2C}, \quad Q = \frac{a^2 E}{k}, \quad C = \frac{a}{k}, \quad W = \frac{a^3 E^2}{2k}, \quad (26)$$

where $W$ is ball charge energy [J]; $C$ is ball capacitance [F]; $Q$ is ball charge [C]. The result of this computation is presented in fig. 7. As can be seen this energy not huge since it is only about 1 - 20 kWh for a ball radius of $a = 5$ m and the electrical intensity is 25 - 100 MV/m. This energy may be restored through ball discharge by emitting the charge into space using a sharp edge.

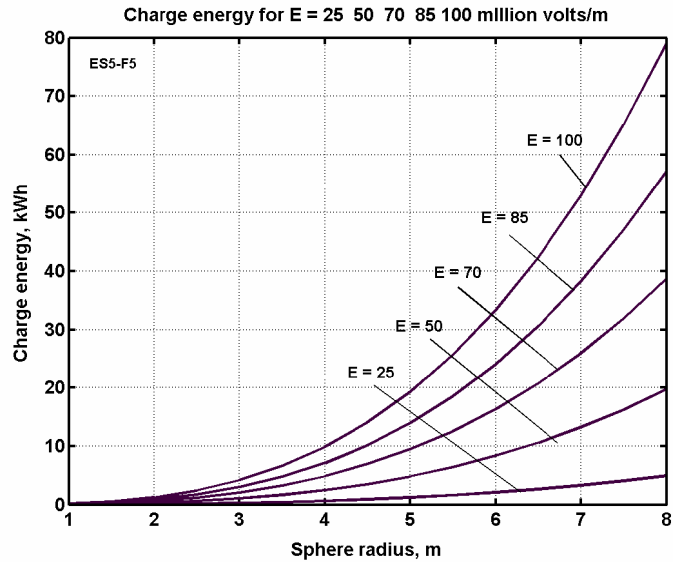

**Fig. 7.** Initial expenditure of electrical energy needed to charge the space apparatus, $a = 1$- 8 m, electrical intensity is 25-100 MV/m and coefficient of efficiency $= 1$.

## Projects

Below the reader will find some examples which highlight the many benefits of the proposed AB-

Ramjet engines. Our parameters are not optimal. Our purpose is simply the demonstration of the potential of AB-Ramjet propulsion.

### Example 1. AB-Ramjets for Earth orbits

**1.** *Brief description of Earth's upper atmosphere.* The Earth's atmosphere consists of 79% nitrogen, 20% oxygen, and 1% other gases. The atmosphere of the Earth may be divided into several distinct layers. The first two are the troposphere (0 - 18 km) and the stratosphere (18 - 90 km). Above the stratosphere is the mesosphere and above that is the *ionosphere* (or *thermosphere*), where many atoms are ionized (gain or lose electrons so they have a net electrical charge). The Sun's ultra-violet radiation and solar wind ionize molecules of the top atmosphere. The ionosphere is very thin, but it is where the aurora takes place, and it is also responsible for absorbing the most energetic photons from the Sun and solar wind. The concentration of ions (= electrons) at day and night time is shown in fig. 8.

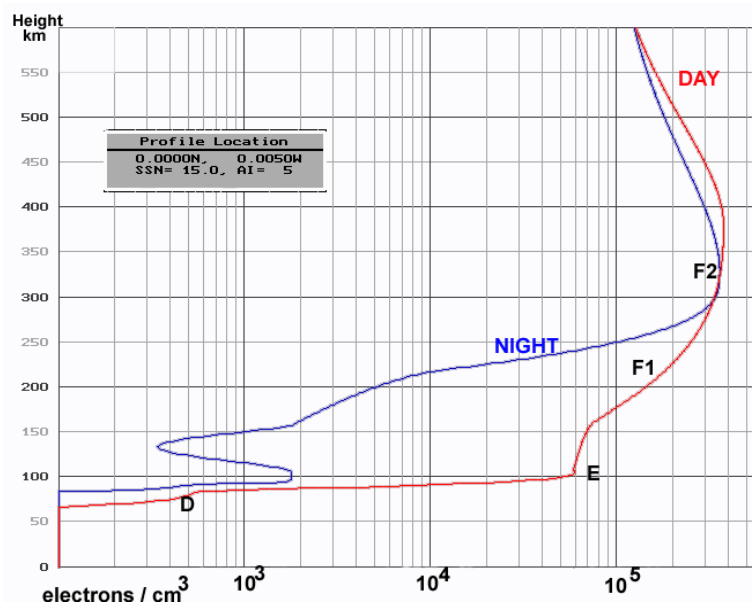

**Fig. 8.** Concentration/cm$^3$ of ions (= electrons) in the day timeand night time in the D, E, F1, and F2 layers of ionosphere.

The ionosphere is divided into the layers D, E, $F_1$, $F_2$. Layer D contains ions of $N_2$ and $O_2$; the layers E, $F_1$, $F_2$ contain ions of $O_2$ and O.

**2.** *Estimations of AB-Ramjet engine data* for low-Earth satellite orbits. Computations are made for an apparatus (AB-Ramjet engine) having speed $V_0$ = 8 km/s, a concentration of $O_2$ ions = $10^5$ ions/cm$^3$, and the electric intensity $E = 10^8$ V/m.

The computation of the AB-radius is presented in fig. 9. An electrostatic collector gathers ions from a surrounding area of more than 10 km$^2$.

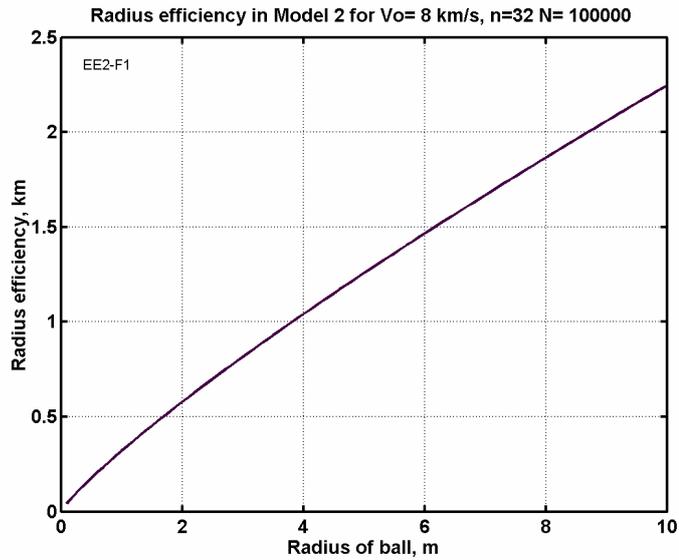

**Fig.9.** AB-radius of charge (radius of efficiency, catch area, clamp area) versus the charge (ball) radius for electric intensity $E$=100 millions V/m, AB-Ramjet engine speed $V_0$ = 8 km/s, $O_2$ ion molar mass $n$ = 32, ion density $N$ = $10^5$ ions/cm$^3$. Equations are [4 - 6].

Produced thrust and requested power are presented in figs. 10, 11. As can be seen, the thrust is significant enough to change the satellite's trajectory and increase the apparatus' speed to a velocity close to that required for an interplanetary trip. The required energy may be obtained from a solar battery.

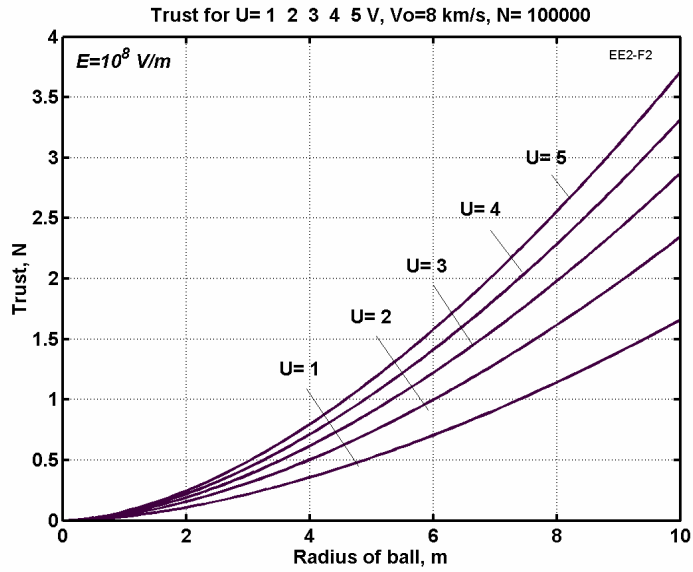

**Fig. 10**. AB-Ramjet engine thrust versus radius of charge (ball) and plate voltage $U$ =1 - 5 V for $V_0$ = 8 km/s, electric intensity $E$=100 millions V/m, and ion density $N$ = $10^5$ ions/cm$^3$. Equation [9].

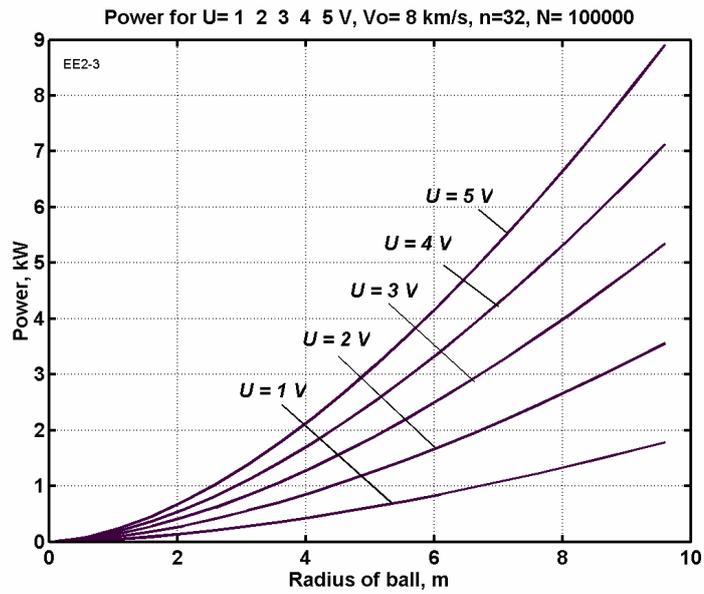

**Fig. 11**. AB-Ramjet engine power versus radius of charge (ball) and plate voltage 1 - 5 for $V_0 = 8$ km/s, electric intensity $E = 100$ millions V/m, and ion density $N = 10^5$ ions/cm$^3$. Equation [11].

Figure 12 shows the drag produced by the AB-Ramjet engine. Figure 13 shows the brake electric energy. This energy may be used by apparatus devices or transferred to other space apparatus.

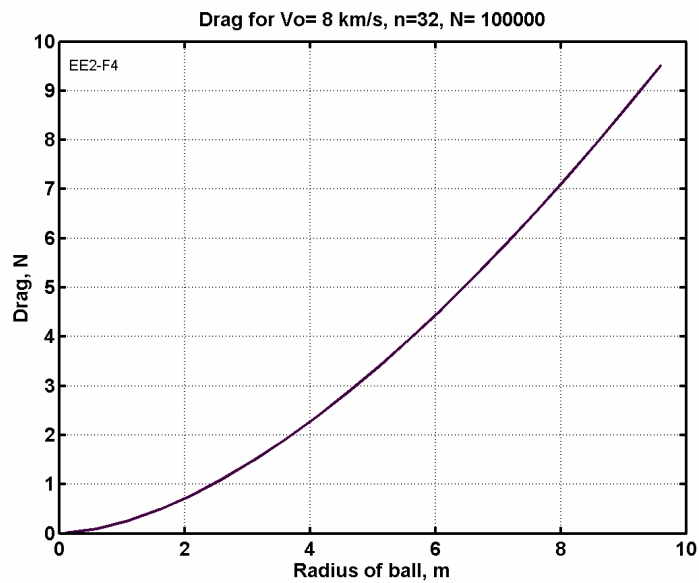

**Fig. 12**. AB-Ramjet engine drag versus radius of charge (ball) for $V_0 = 8$ km/s, electric intensity $E = 100$ millions V/m, and ion density $N = 10^5$ ions/cm$^3$. Equation [9a].

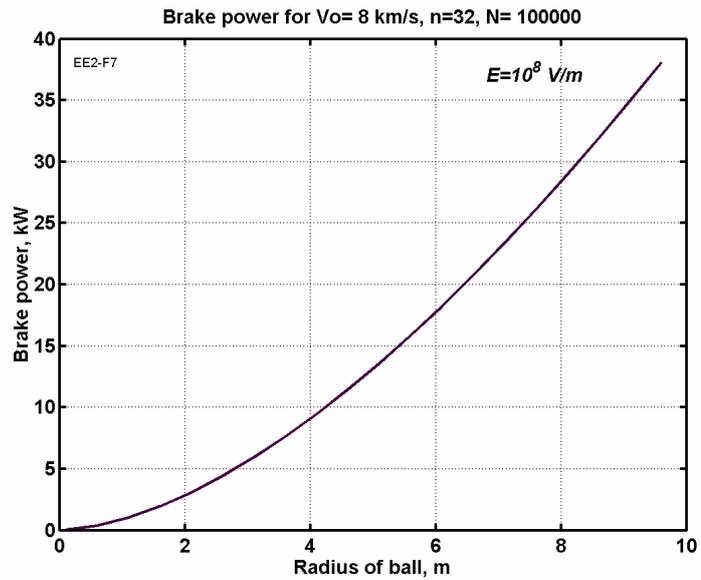

**Fig. 13**. AB-Ramjet engine power versus radius of charge (ball) for $V_0 = 8$ km/s, electric intensity $E = 100$ millions V/m, and ion density $N = 10^5$ ions/cm$^3$. Equation [11].

The offered AB-engine may be used as an accelerator or brake for interplanetary space apparatus. The method of acceleration is shown in fig. 14. In the acceleration regime in region 7, the thruster of the AB-engine accelerates a probe and increases the apogee of the elliptic trajectory. In the final trajectory, the probe is separated from the engine, gets a small impulse, and flies away into space. When the probe returns from space flight, the sequence is reversed and the AB-engine works as brake.

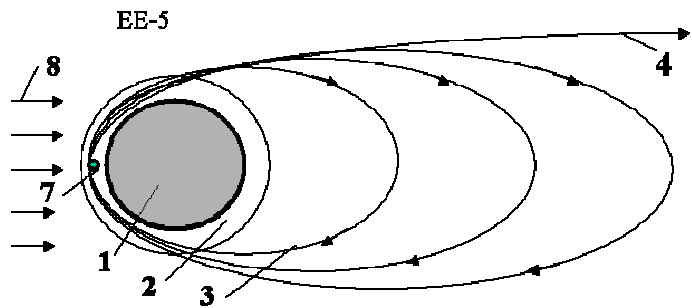

**Fig.14.** Using the AB-engine as an accelerator and a brake for interplanetary flight. Notation: 1 - Earth, 2 - Earth's atmosphere, 3 - accelerator trajectories, 4 - final trajectory of interplanetary probe, 7 - region of acceleration and brake, 8 - solar light.

The particle mass flow and the electric current flow (computed by equation [8],[9]) are shown in figs. 15, 16. The electric current is significantly large, creating a magnetic field which helps to straighten the particle's trajectory.

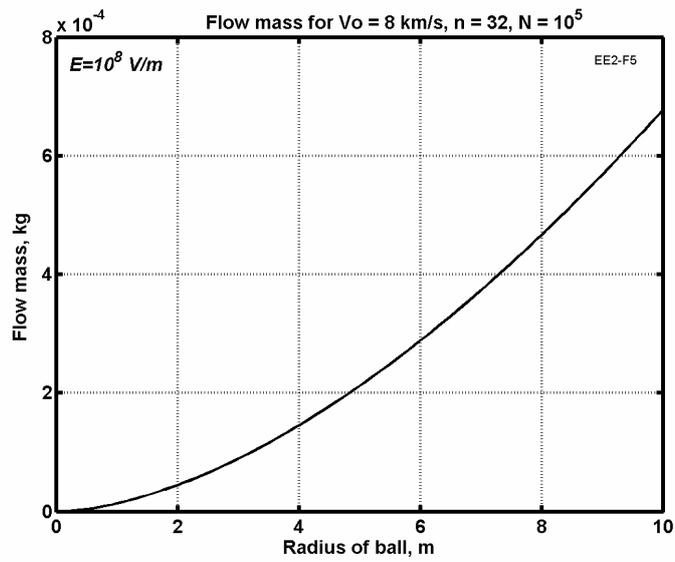

**Fig. 15.** Ion mass flow through the AB-engine. Equation [15].

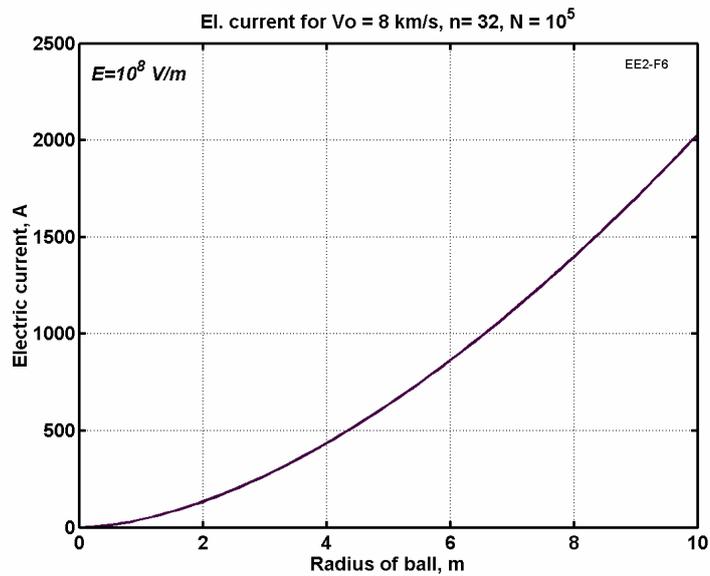

**Fig. 16.** Electric current flow through the AB-engine. Equation [16].

The additional speed gained by the ions in AB-engine as computed by equation [7] via the interplate voltage is shown in figure 17. Acceleration is produced in the thrust regime and deceleration is produced in the brake regime.

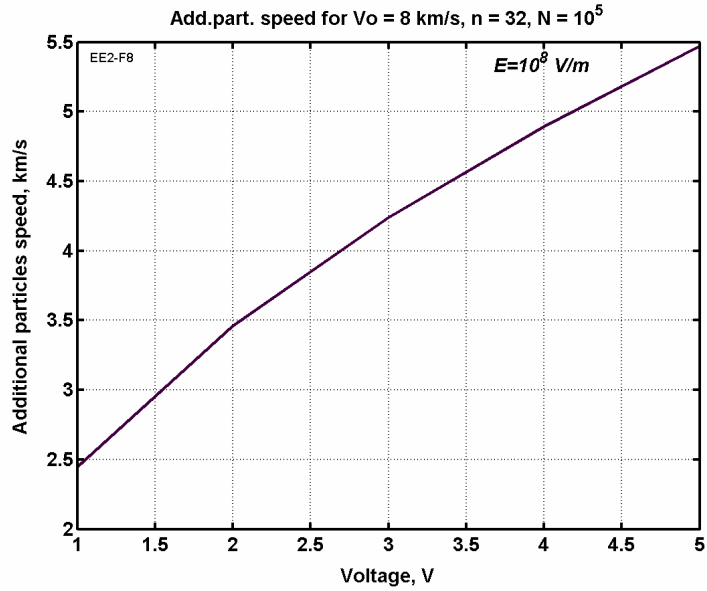

**Fig. 17**. The additional speed getting by ions when they move between engine plates.

The size of the plates computed by equation [2b] is shown in fig. 18. As can be seen, the size is very small and the plates can be located inside of the cylindrical charged core (fig. 2f).

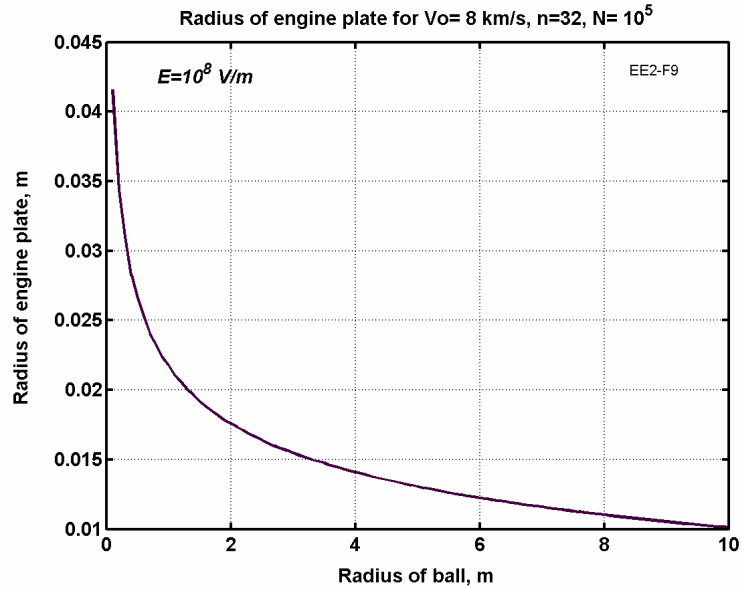

**Fig.18**. Size of AB-engine plates versus the ball radius. Equation [2b].

# Example 2. Interplanetary AB-Ramjet.

**Brief information about the solar wind**. The Sun emits plasma which is a continuous outward flow (solar wind) of ionized solar gas throughout our solar system. The solar wind contains about 90% protons and electrons and some quantities of ionized $\alpha$-particles and gases. It attains speeds in the range of 300-750 km/s and has a flow density of $5\times10^7$ - $5\times10^8$ protons/ electrons/cm$^2$s. The observed speed rises systematically from low values (300-400 km/s) to high values (650-700 km/s) in 1 or 2 days and then returns to low values during the next 3 to 5 days (Fig. 19). Each of these high-speed streams tends to appear at approximately 27-day intervals or to recur with the rotation period of the Sun. On days of high Sun activity the solar wind speed reaches 1000 (and more) km/s and its flow density is $10^9$ - $10^{10}$ protons/electrons/ cm$^2$s with 8 -70 particles per cm$^3$. The Sun has high activity periods several days each year.

The pressure of the solar wind is very small. For full braking it is in the interval $2.5 \times 10^{-10} \div 6.3 \times 10^{-9}$ N/m$^2$. This value is doubled when the particles have full reflection. The interstellar medium also has high-energy particles. Their density is about 1 particle/cm$^3$.

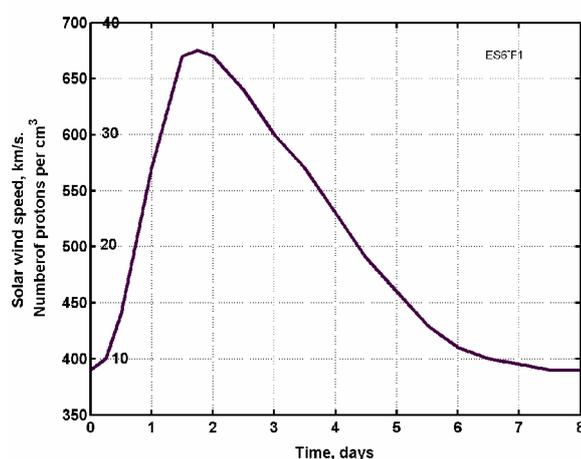

**Fig. 19**. Speed and density variations in solar wind. The speed is in km/s, the density is in protons/cm$^3$.

**Estimation of main parameters of the AB-engine**. Let us estimate the main parameters of the AB-Ramjet engine for interplanetary apparatus. Interplanetary apparatus using a charged ball for solar wind drag was considered by the author in [54]. This work shows a great potential for these apparatus. The suggested AB-Ramjet engine (fig. 2) is different from the propulsion engine in [54]. The AB-Ramjet has plates which can produce thrust, work as a generator of electrical energy, create stronger drag, and improve the control of the value and direction of both the drag and thrust. The central charge has a cylindrical form which dramatically decreases the discharging of the central charged core. The proposed AB-Ramjet can simultaneously produce useful drag (or thrust) and electric energy. It may seem astonishing, but an AB-Ramjet located in the strong solar wind would be capable of achieving a speed between 400 - 750 km/s. If we install the wind engine in a conventional space ship, the engine will produce sufficient energy and drag which would also useful if space ship moved in the wind direction. The drag produced by solar wind is a useful thrust for moving from Earth to outer Earth orbit: Saturn - Pluto. It may appear that this apparatus is lacking because it can only move away from the Sun. However, that is not the case. The AB-ramjet apparatus can decrease the orbit speed and the Sun's gravity will move it back to Earth orbit.

Estimates of the main parameters of the AB-ramjet interplanetary engine and apparatus are given below. The equations used in computing the estimates are in theoretical section [Eqs. 4 -6]. Figure 20 shows the AB-radius of the efficiency area (catch area, clamp area) versus the charge radius of ball. Data used for computation: electric intensity $E$=100 millions V/m, solar wind speed $V_0$ = 400 km/s, solar wind density $N$ = 10 protons/cm$^3$.

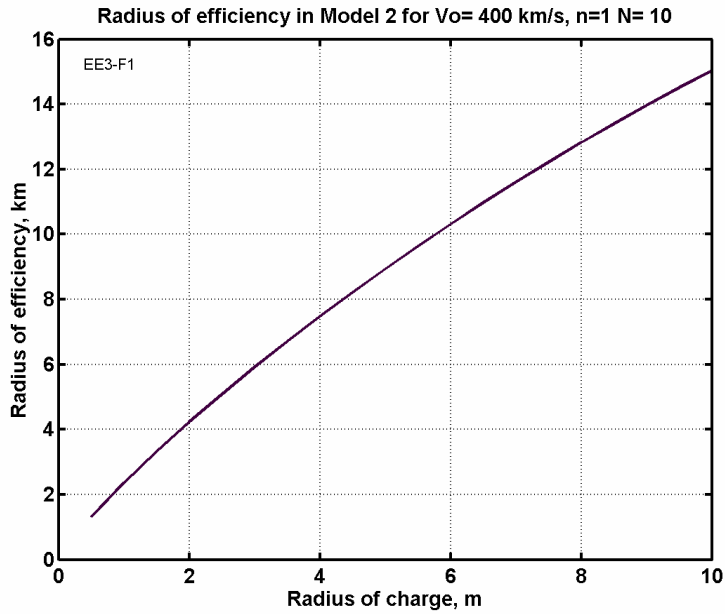

**Fig. 20.** Radius of efficiency (catch area, clamp area) versus the charge radius for electric intensity $E$=100 millions V/m, solar wind speed $V_0$ = 400 km/s, solar wind density $N$ = 10 protons/cm$^3$. Equations are [4 - 6].

Mass and charge flow (electric current) through the AB-engine were computed using equations [8] and [9]. Results are presented in figs. 21 and 22 for the following conditions: electric intensity $E$=100 millions V/m, solar wind speed $V_0$ = 400 km/s, solar wind density $N$ = 10 protons/cm$^3$. The flow mass is small (5 milligram/s), however the electric current is large.

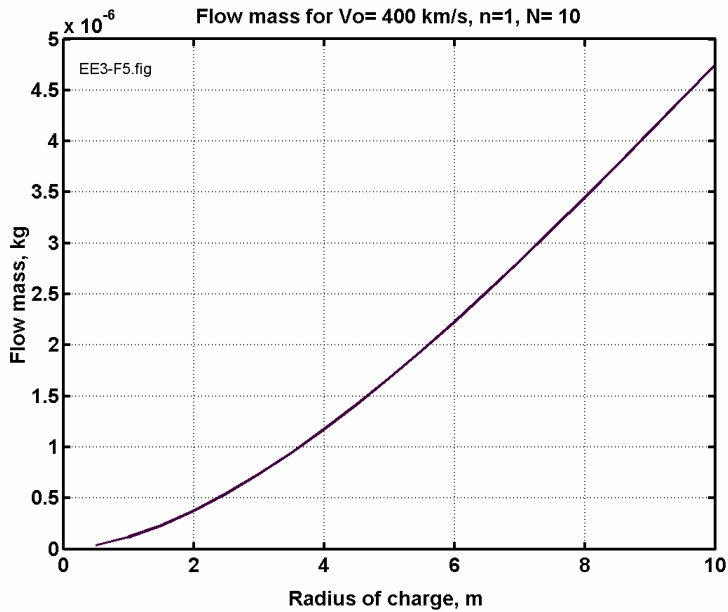

**Fig.21.** Flow of particles mass through the AB-Ramjet engine versus the charge radius for electric intensity $E$ = 100 millions V/m, solar wind speed $V_0$ = 400 km/s, solar wind density $N$ = 10 protons/cm$^3$.

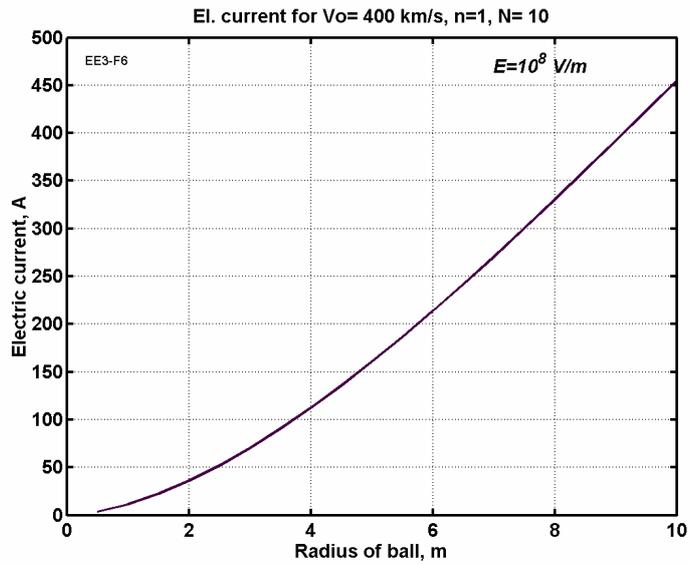

**Fig.22**. Flow electric currency through the AB-Ramjet engine versus the charge radius for electric intensity $E = 100$ millions V/m, solar wind speed $V_0 = 400$ km/s, solar wind density $N = 10$ protons/cm$^3$.

Figure 23 shows the maximal drag for full reflection versus the charge radius computed via equation 9a for the conditions: electric intensity $E=100$ millions V/m, solar wind speed $V_0 = 400$ km/s, solar wind density $N = 10$ protons/cm$^3$. As can be seen, the useful drag is sufficiently large and can produce significant acceleration for space apparatus. More detailed computations for the conventional charge are presented in [54].

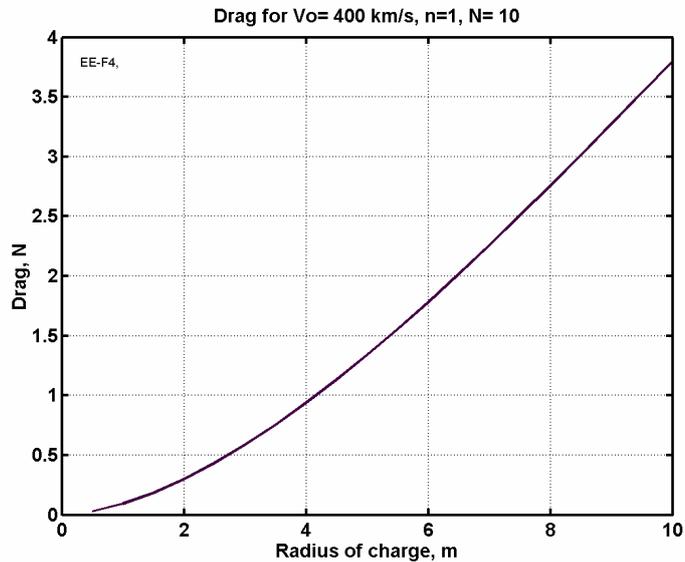

**Fig. 23.** Maximal drag versus the charge radius for electric intensity $E=100$ millions V/m, solar wind speed $V_0 = 400$ km/s, solar wind density $N = 10$ protons/cm$^3$. Full reflection.

Thrust versus voltage between plates is shown in fig. 24 and the required energy for it is presented in fig. 25. For a thrust of 0.2 N, the required power is 4.5 kW. We can increase the voltage and obtain more thrust but the needed energy may not be acceptable for the apparatus.

If we change polarity of the plates, we get the same energy (fig. 26) from apparatus braking, plus additional drag.

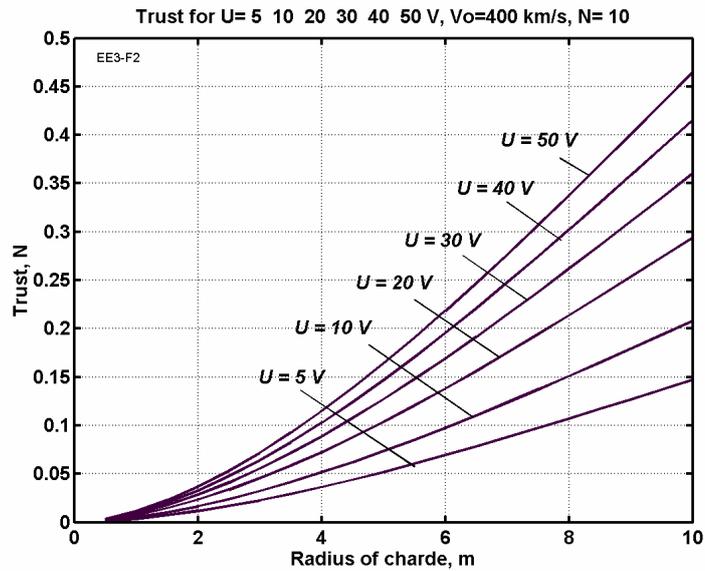

**Fig.24.** AB-Ramjet engine thrust versus radius of charge and plate voltage for $V_0 = 400$ km/s and flow density 10 protons/cm$^3$.

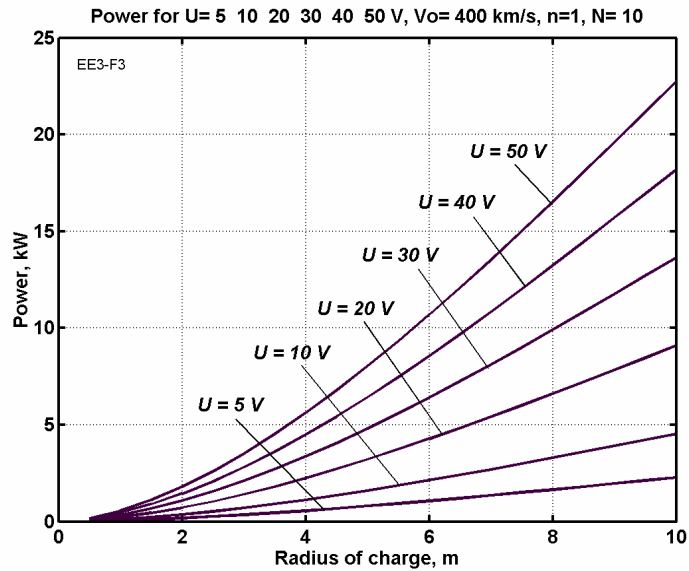

**Fig.25.** Requested AB-Ramjet engine power for thrust versus radius of charge and plate voltage for $V_0 = 400$ km/s and flow density 10 protons/cm$^3$.

 The maximum energy and half of full reflection drag will be obtained when the positive value of $(V_0 - V) > 0$ is close to zero. Here $V_0$ is solar wind speed and $V$ is apparatus speed (positive direction from Sun). Using equation [7] ($mV^2/2 = eU$), for a wind speed $V_0 = 400$ km/s, the calculation gives $U = 835$ V. This means for a highly charged ball, $U = 0$ produces good reflection of the protons, near maximum drag, and zero positive electric energy. A $U$ slightly less than 835 V produces near maximum energy and a drag close to half of the maximum drag (fig.23). The computation for this case is presented in fig. 26. The power is significantly large and the engine may be used for both the apparatus and the charging of the electrostatic collector (core).

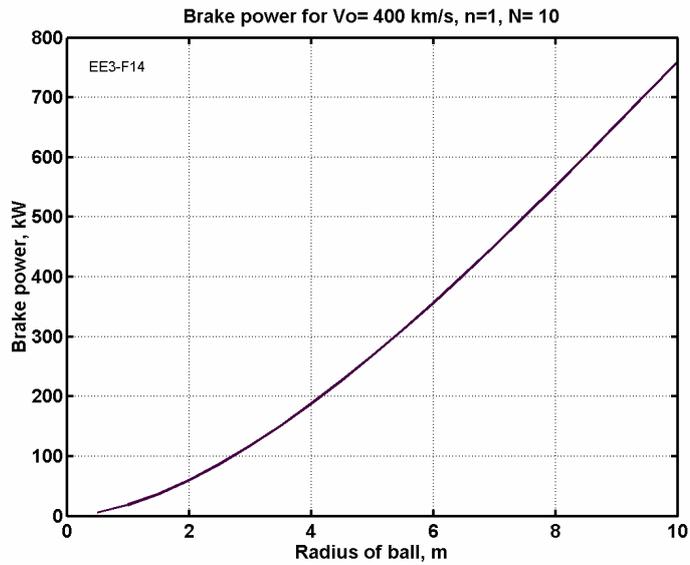

**Fig. 26.** The maximum electric energy which can be obtained from the AB-Ramjet engine versus charge radius. In this case the drag (half of full drag) accelerates the apparatus to far planets.

Let us find the size of the electrostatic collector. The maximal size of the plate radius is between the parameter $R_h$ and the minimal radius of the hyperbolic trajectory. These magnitudes are presented for the outermost particles in fig. 27. The computation shows a particle flow that has maximum density in the plate center and small density in the plate ends.

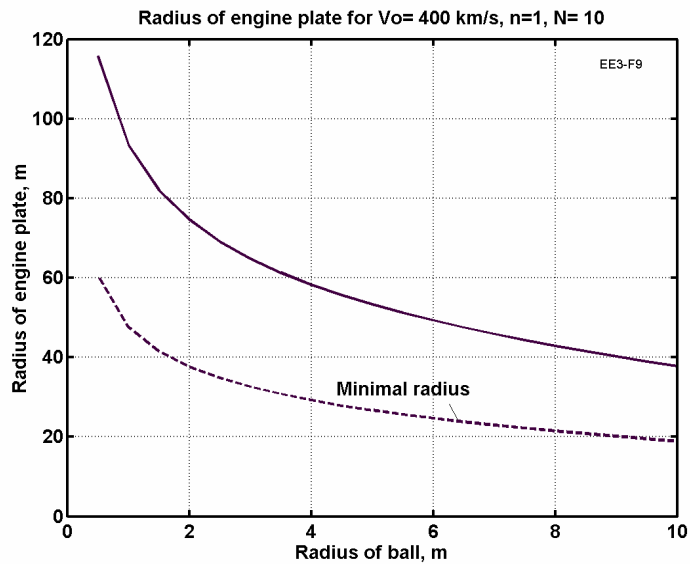

**Fig.27.** The minimal radius and parameter $R_h$ of hyperbolic trajectory of outermost particles.

We can determine a plate radius by the lower curve. For a charge (ball) radius of 10 meter, a plate radius of 20 meters is sufficient.

Let us estimate the discharging power for a radius of central charge of 10 meters. Let the wire diameter of the engine plate net be 0.1 mm with a cell size of 10×10 cm. The light transparency coefficient is $c_1 = 4 \times (0.001)^2 = 4 \times 10^{-6}$ (Eq. [19]). The brake coefficient of wire is $c_2 = 10^{-5}$ (Eq. [21]). The total transparency coefficient is $c_3 = c_1 \times c_2 = 4 \times 10^{-11}$ (Eq. [22]). Loss of energy in the plate is $L = c_3 IU = 4 \times 10^{-11} \times 450 \times 10^9 = 1.8$ W. If we take into consideration the cylindrical central charge and other support elements and increase this value by 10, 100, 1000 times, the brake power (750 kW, fig. 26) will be enough for the compensation this energy.

The maximum electric force between plates for a distance of 2 meters and a voltage of 1000 V is 0.14 N (eq. [18]). This means the support elements can be very light. The mass of the two quadratic plates of size 20×20 m having aluminum wires is 1.7 kg.

## Conclusion

The primary research and computations of the suggested AB-engine show the numerous possibilities and perspectives of the space AB-ramjet engines. The density of the charged space particles is very small. But the proposed electrostatic collector can effectively gather the particles from a huge surrounding area and accelerate or brake them, generating thrust or braking on the order of several Newtons. The high speed solar wind allows simultaneously obtainment of useful drag (thrust) and great electrical energy. The simplest electrostatic gatherer accelerates a 100 kg probe up to a velocity of 100 km/s [54]. The probe offers flights into Mars orbit of about 70 days, to Jupiter orbit in about 150 days, to Saturn orbit in about 250 days, to Uranus orbit in about 450 days, to Neptune orbit in about 650 days, and to Pluto orbit in about 850 days.

The suggested electric gun is simple and can transfer energy (charge by electron beam) over a long distance to other space apparatus.

The author has developed the initial theory and the initial computations to show the possibility of the offered concepts. He calls on scientists, engineers, space organizations, and companies to research and develop the proposed perspective concepts.

## References
(Some of these works the reader finds in author side: http://Bolonkin.narod.ru/p65.htm)

1. Bolonkin, A.A., (1965a), "Theory of Flight Vehicles with Control Radial Force". Collection *Researches of Flight Dynamics*, Mashinostroenie Publisher, Moscow, , pp. 79-118, 1965, (in Russian). Intern.Aerospace Abstract A66-23338#(Eng).
2. Bolonkin A.A., (1965c), Optimization of Trajectories of Multistage Rockets. Collection *Researches of Flight Dynamics*. Moscow, 1965, p. 20 -78 (in Russian). International Aerospace Abstract A66-23337# (English).
3. Bolonkin, A.A., (1982a), Installation for Open Electrostatic Field, Russian patent application #3467270/21  116676, 9 July, 1982 (in Russian), Russian PTO.
4. Bolonkin, A.A., (1982b), Radioisotope Propulsion. Russian patent application #3467762/25  116952, 9 July 1982 (in Russian), Russian PTO.
5. Bolonkin, A.A., (1982c), Radioisotope Electric Generator. Russian patent application #3469511/25  116927. 9 July 1982 (in Russian), Russian PTO.
6. Bolonkin, A.A., (1983a), Space Propulsion Using Solar Wing and Installation for It, Russian patent application #3635955/23  126453, 19 August, 1983 (in Russian), Russian PTO.
7. Bolonkin, A.A., (1983b), Getting of Electric Energy from Space and Installation for It, Russian patent application #3638699/25  126303, 19 August, 1983 (in Russian), Russian PTO.
8. Bolonkin, A.A., (1983c), Protection from Charged Particles in Space and Installation for It, Russian patent application #3644168  136270, 23 September 1983, (in Russian), Russian PTO.
9. Bolonkin, A. A., (1983d), Method of Transformation of Plasma Energy in Electric Current and Installation for It. Russian patent application #3647344  136681 of 27 July 1983 (in Russian), Russian PTO.
10. Bolonkin, A. A., (1983e), Method of Propulsion using Radioisotope Energy and Installation for It.  of Plasma Energy in Electric Current and Installation for it. Russian patent application #3601164/25  086973  of 6 June, 1983 (in Russian), Russian PTO.
11. Bolonkin, A. A.,(1983f),  Transformation of Energy of Rarefaction Plasma in Electric Current and Installation for it. Russian patent application #3663911/25  159775, 23 November 1983 (in Russian), Russian PTO.

12. Bolonkin, A. A., (1983g), Method of a Keeping of a Neutral Plasma and Installation for it. Russian patent application #3600272/25 086993, 6 June 1983 (in Russian), Russian PTO.
13. Bolonkin, A.A.,(1983h), Radioisotope Electric Generator. Russian patent application #3620051/25 108943, 13 July 1983 (in Russian), Russian PTO.
14. Bolonkin, A.A., (1983i), Method of Energy Transformation of Radioisotope Matter in Electricity and Installation for it. Russian patent application #3647343/25 136692, 27 July 1983 (in Russian), Russian PTO.
15. Bolonkin, A.A., (1983j), Method of stretching of thin film. Russian patent application #3646689/10 138085, 28 September 1983 (in Russian), Russian PTO.
16. Bolonkin, A.A., (1987), "New Way of Thrust and Generation of Electrical Energy in Space". Report ESTI, 1987, (Soviet Classified Projects).
17. Bolonkin, A.A., (1990), "Aviation, Motor and Space Designs", Collection *Emerging Technology in the Soviet Union*, 1990, Delphic Ass., Inc., pp.32-80 (English).
18. Bolonkin, A.A., (1991), *The Development of Soviet Rocket Engines*, 1991, Delphic Ass.Inc.,122 p. Washington, (in English).
19. Bolonkin, A.A., (1992a), "A Space Motor Using Solar Wind Energy (Magnetic Particle Sail)". The World Space Congress, Washington, DC, USA, 28 Aug. - 5 Sept., 1992, IAF-0615.
20. Bolonkin, A.A., (1992b), "Space Electric Generator, run by Solar Wing". The World Space Congress, Washington, DC, USA, 28 Aug. -5 Sept. 1992, IAF-92-0604.
21. Bolonkin, A.A., (1992c), "Simple Space Nuclear Reactor Motors and Electric Generators Running on Radioactive Substances", The World Space Congress, Washington, DC, USA, 28 Aug. - 5 Sept., 1992, IAF-92-0573.
22. Bolonkin, A.A. (1994), "The Simplest Space Electric Generator and Motor with Control Energy and Thrust", 45th International Astronautical Congress, Jerusalem, Israel, 9-14 Oct., 1994, IAF-94-R.1.368 .
23. Bolonkin, A.A., (2002a), "Non-Rocket Space Rope Launcher for People", IAC-02-V.P.06, 53rd International Astronautical Congress, The World Space Congress - 2002, 10-19 Oct 2002, Houston, Texas, USA.
24. Bolonkin, A.A,(2002b), "Non-Rocket Missile Rope Launcher", IAC-02-IAA.S.P.14, 53rd International Astronautical Congress, The World Space Congress - 2002, 10-19 Oct 2002, Houston, Texas, USA.
25. Bolonkin, A.A.,(2002c), "Inexpensive Cable Space Launcher of High Capability", IAC-02-V.P.07, 53rd International Astronautical Congress, The World Space Congress - 2002, 10-19 Oct 2002, Houston, Texas, USA.
26. Bolonkin, A.A.,(2002d), "Hypersonic Launch System of Capability up 500 tons per day and Delivery Cost $1 per Lb". IAC-02-S.P.15, 53rd International Astronautical Congress, The World Space Congress - 2002, 10-19 Oct 2002, Houston, Texas, USA.
27. Bolonkin, A.A.,(2002e), "Employment Asteroids for Movement of Space Ship and Probes". IAC-02-S.6.04, 53rd International Astronautical Congress, The World Space Congress - 2002, 10-19 Oct 2002, Houston, Texas, USA.
28. Bolonkin, A.A., (2002f), "Optimal Inflatable Space Towers of High Height". COSPAR-02 C1.1-0035-02, 34th Scientific Assembly of the Committee on Space Research (COSPAR), The World Space Congress - 2002, 10-19 Oct 2002, Houston, Texas, USA.
29. Bolonkin, A.A., (2002g), "Non-Rocket Earth-Moon Transport System", COSPAR-02 B0.3-F3.3-0032-02, 02-A-02226, 34th Scientific Assembly of the Committee on Space Research (COSPAR), The World Space Congress - 2002, 10-19 Oct 2002, Houston, Texas, USA.
30. Bolonkin, A. A.,(2002h) "Non-Rocket Earth-Mars Transport System", COSPAR-02 B0.4-C3.4-0036-02, 34th Scientific Assembly of the Committee on Space Research (COSPAR), The World Space Congress - 2002, 10-19 Oct 2002, Houston, Texas, USA.
31. Bolonkin, A.A.,(2002i). "Transport System for Delivery Tourists at Altitude 140 km". IAC-02-IAA.1.3.03, 53rd International Astronautical Congress, The World Space Congress - 2002, 10-19 Oct. 2002, Houston, Texas, USA.
32. Bolonkin, A.A., (2002j), "Hypersonic Gas-Rocket Launch System." AIAA-2002-3927, 38th AIAA/ASME/SAE/ASEE Joint Propulsion Conference and Exhibit, 7-10 July 2002. Indianapolis, IN, USA.

33  Bolonkin, A.A., (2003a), "Air Cable Transport", *Journal of Aircraft*, Vol. 40, No. 2, March-April 2003.
34  Bolonkin, A.A., (2003b), "Optimal Inflatable Space Towers with 3-100 km Height", *JBIS*, Vol. 56, No 3/4, pp. 87-97, 2003.
35  Bolonkin, A.A.,(2003c), "Asteroids as Propulsion Systems of Space Ships", *JBIS*, Vol. 56, No 3/4, pp. 97-107, 2003.
36  Bolonkin A.A., (2003d), "Non-Rocket Transportation System for Space Travel", *JBIS*, Vol. 56, No 7/8, pp. 231-249, 2003.
37  Bolonkin A.A., (2003e), "Hypersonic Space Launcher of High Capability", *Actual problems of aviation and aerospace systems*, Kazan, No. 1(15), Vol. 8, 2003, pp. 45-58.
38  Bolonkin A.A., (2003f), "Centrifugal Keeper for Space Stations and Satellites", *JBIS*, Vol. 56, No 9/10, pp. 314-327, 2003.
39  Bolonkin A.A., (2003g), "Non-Rocket Earth-Moon Transport System", *Advances in Space Research*, Vol. 31/11, pp. 2485-2490, 2003, Elsevier.
40  Bolonkin A.A., (2003h), "Earth Accelerator for Space Ships and Missiles". *JBIS*, Vol. 56, No. 11/12, 2003, pp. 394-404.
41  Bolonkin A.A., (2003i), "Air Cable Transport and Bridges", TN 7567, International Air & Space Symposium - The Next 100 Years, 14-17 July 2003, Dayton, Ohio, USA.
42  Bolonkin, A.A., (2003j), "Air Cable Transport System", *Journal of Aircraft*, Vol. 40, No. 2, March-April 2003, pp. 265-269.
43  Bolonkin A.A.,(2004a), "Kinetic Space Towers and Launchers ', *JBIS*, Vol. 57, No 1/2, pp. 33-39, 2004.
44  Bolonkin A.A.,(2004b), "Optimal trajectory of air vehicles", *Aircraft Engineering and Space Technology*, Vol. 76, No. 2, 2004, pp. 193-214.
45  Bolonkin A.A., (2004c), "Long Distance Transfer of Mechanical Energy", International Energy Conversion Engineering Conference at Providence RI, Aug. 16-19, 2004, AIAA-2004-5660.
46  Bolonkin, A.A., (2004d), "Light Multi-Reflex Engine", Journal *JBIS*, Vol. 57, No 9/10, pp. 353-359, 2004.
47  Bolonkin, A.A., (2004e), "Kinetic Space Towers and Launchers", Journal *JBIS*, Vol. 57, No 1/2, pp. 33-39, 2004.
48  Bolonkin, A.A., (2004f), "Optimal trajectory of air and space vehicles", *AEAT*, No 2, pp. 193-214, 2004.
49  Bolonkin, A.A.,(2004g), "Hypersonic Gas-Rocket Launcher of High Capacity", Journal *JBIS*, Vol. 57, No 5/6, pp. 167-172, 2004.
50  Bolonkin, A.A., (2004h), "High Efficiency Transfer of Mechanical Energy". International Energy Conversion Engineering Conference at Providence RI, USA. 16-19 August, 2004, AIAA-2004-5660.
51  Bolonkin, A.A., (2004i), "Multi-Reflex Propulsion System for Space and Air Vehicles", *JBIS*, Vol. 57, No 11/12, 2004, pp. 379-390.
52  Bolonkin A.A.,(2005a) "High Speed Catapult Aviation", AIAA-2005-6221, Atmospheric Flight Mechanic Conference - 2005, 15-18 August, 2005, USA.
53  Bolonkin A.A., (2005a), Electrostatic Solar Wind Propulsion System, AIAA-2005-3653. 41-st Propulsion Conference, 10-12 July, 2005, Tucson, Arizona, USA.
54  Bolonkin A.A., (2005b), Electrostatic Utilization of Asteroids for Space Flight, AIAA-2005-4032. 41 Propulsion Conference, 10-12 July, 2005, Tucson, Arizona, USA.
55  Bolonkin A.A., (2005c), Kinetic Anti-Gravitator, AIAA-2005-4504. 41-st Propulsion Conference, 10-12 July, 2005, Tucson, Arizona, USA.
56  Bolonkin A.A., (2005d), Sling Rotary Space Launcher, AIAA-2005-4035. 41-st Propulsion Conference, 10-12 July, 2005, Tucson, Arizona, USA.
57  Bolonkin A.A., (2005e), Radioisotope Space Sail and Electric Generator, AIAA-2005-4225. 41-st Propulsion Conference, 10-12 July, 2005, Tucson, Arizona, USA.
58  Bolonkin A.A., (2005f), Guided Solar Sail and Electric Generator, AIAA-2005-3857. 41-st Propulsion Conference, 10-12 July, 2005, Tucson, Arizona, USA.
59  Bolonkin A.A., (2005g), Problems of Electrostatic Levitation and Artificial Gravity, AIAA-2005-4465. 41 Propulsion Conference, 10-12 July, 2005, Tucson, Arizona, USA.
60  Bolonkin A.A., (2006), *Non-Rocket Space Launch and Flight*, Elsevier, London, pp. 488.